\documentclass[sigconf]{acmart}

\usepackage{subfigure}
\usepackage[linesnumbered,ruled,vlined]{algorithm2e}
\SetKwInput{KwInput}{Input}                
\SetKwInput{KwOutput}{Output}              

\AtBeginDocument{%
  \providecommand\BibTeX{{%
    \normalfont B\kern-0.5em{\scshape i\kern-0.25em b}\kern-0.8em\TeX}}}

\acmYear{ }
\acmDOI{}
\acmConference[AdKDD'22]{AdKDD 2022 Workshop}{August 15, 2022}{Washington, DC, USA}
\acmPrice{}
\acmISBN{}
\setcopyright{none} 
\begin{document}

\title{Online Meta-Learning for Model Update Aggregation in Federated Learning for Click-Through Rate Prediction}

\author{Xianghang Liu}
\email{xianghang.liu@huawei.com}
\affiliation{%
  \institution{Huawei London Research Centre}
  \city{London}
  \country{UK}}

\author{Bartłomiej Twardowski}
\email{bartlomiej.twardowski@huawei.com}
\affiliation{%
  \institution{Huawei Ireland Research Centre}
  \city{Dublin}
  \country{Ireland}}

\author{Tri Kurniawan Wijaya}
\email{tri.kurniawan.wijaya@huawei.com}
\affiliation{%
  \institution{Huawei Ireland Research Centre}
  \city{Dublin}
  \country{Ireland}}

\renewcommand{\shortauthors}{X.Liu, B.Twardowski, T.K.Wijaya}

\newcommand{\minisection}[1]{\noindent {\bf #1}}

\begin{abstract}
In Federated Learning (FL) of click-through rate (CTR) prediction, users' data is not shared for privacy protection. The learning is performed by training locally on client devices and communicating only model changes to the server. 
There are two main challenges:
(i) the client heterogeneity, making FL algorithms that use the weighted averaging to aggregate model updates from the clients have slow progress and unsatisfactory learning results; and (ii) the difficulty of tuning the server learning rate with trial-and-error methodology due to the big computation time and resources needed for each experiment. To address these challenges, we propose a simple online meta-learning method to learn a strategy of aggregating the model updates, which adaptively weighs the importance of the clients based on their attributes and adjust the step sizes of the update. We perform extensive evaluations on public datasets. Our method significantly outperforms the state-of-the-art in both the speed of convergence and the quality of the final learning results.
\end{abstract}


\maketitle

\section{Introduction}
Federated Learning (FL) \cite{mcmahan2017communication} is a machine learning paradigm where user data is not shared for privacy protection.
The model is trained via arranged communications between the server and many system clients. 
In each communication round, the selected clients run local optimization algorithms for multiple steps and send the local model updates back to the server.
The server aggregates the received local updates and then applies a gradient-based server optimizer, taking the aggregated update as the gradient input.

In recent years, deep neural network based models have proven to be successful in click-through rate (CTR) prediction~\cite{Song2019autoint, wang2021dcn}. 
Based on user, item and context features, they predict CTR without tedious crafting of input features or higher-order interactions between them. 

In this work, we focus on FL application for CTR prediction models.
Specifically, we propose an algorithm to learn the aggregation strategy of the local model updates in FL CTR models, 
where the aggregation strategy is adjusted according to the feedback during the learning process.
There are two main motivations for our work described below. 

\minisection{Weighting of local updates and client heterogeneity.}
Weighted averaging is a commonly-used strategy for local update aggregation in FL.
When all clients are homogeneous, taking the average over local updates gives good results \cite{li2019convergence}.
In practice, however, the clients are often heterogeneous.
This client heterogeneity is a major challenge in FL for CTR prediction models and recommender systems (RecSys) in general, because it causes inconsistencies between the client and the global objectives; 
it slows down the training progress and often leads to suboptimal final results.
This issue is also known as \textit{client drift} \cite{karimireddy20scaffold}.

To obtain faster learning progress and better final solutions, a good client weighting strategy for update aggregation should be 
(1) aware of the client heterogeneity: the importance weight of a client should depend on its attributes;
(2) adaptive: at different phases of the learning process, the strategy may focus on different sets of clients; 
(3) parameter-wise: not all parameters behave in the same way during a training process, the strategy should support different learning dynamics for them. 
For example, in FL of deep CTR models, there are parameters for sparse feature embeddings that are only updated by a small set of clients. On the contrary, dense fully-connected layers are updated by all clients in each round. The optimal weighting strategies for these two sets of parameters are likely to be different.

\minisection{Server learning rate tuning.}
Learning rate is arguably one of the most important hyper-parameters for gradient-based learning algorithms in centralised machine learning \cite{Bengio2012}. 
Likewise, server learning rate, used by the server optimizer to control the step-size of the aggregated update, has an important role in FL~\cite{charles2020outsized, reddi2021adaptive, wang2021field}. 
Similar to the case of client weighting, it is desirable that the server learning rate schedule is \textit{adaptive} and \textit{parameter-wise}.

In this paper, we propose an online meta-learning method to learn a strategy to aggregate the
local model updates in federated learning of CTR prediction models, which has the advantages
of being aware of client heterogeneity, adaptive to learning process, and parameter-wise.
Our contributions include:
\begin{itemize}
    \setlength\itemsep{0em}
    \item Meta-learning formulation of the aggregation problem in FL to 
    compute the gradients of meta-parameters without the need of an extra dataset.
    \item A new online meta-learning method (MetaUA) that jointly learns the target CTR prediction model and the meta-aggregation model. 
    \item Extensive experimental evaluations, including: comparison to other methods on 
    four publicly available datasets, ablation study, analysis of clients' attributes selection, 
    and method robustness to varying percentages of participating clients and meta-learning rate. MetaUA 
    outperforms state-of-the-art methods in both the speed of convergence and the final results in few different
    settings.
\end{itemize}

\vspace{-2ex}
\section{Related work}
The first FL algorithm FedAvg was proposed ~\cite{mcmahan2017communication}.    
Since then, there have been many FL works dedicated to RecSys \cite{fedfast, flanagan2020fedmvmf, lin2020metamf}. 
While most of these works focus on learning generalized matrix factorization models \cite{He2017ncf}, 
our work is agnostic to model architectures. 

Client heterogeneity is recognized as one of the key challenges in FL~\cite{wang2021field, Kairouz2021Advances}.
A few approaches have been proposed to address the client drift problem caused by heterogeneity:
FedNova~\cite{Wang2020fednova} normalizes the local updates before averaging to eliminate the objective inconsistency; 
SCAFFOLD~\cite{karimireddy20scaffold} uses control variates to correct the drift in the local updates;  
FedProx~\cite{Li2020Fedprox} adds a proximal term in the local objectives to stabilize the learning.

On the aggregation of local model updates, FedAvg uses a weighted average by the number of samples received from the clients. 
Inspired by the attention mechanism \cite{Vaswani2017Attention, Ji2019FedAtt, fedfast} use the Euclidean distance between local and global models as the importance weights of clients. 
In a similar spirit, \cite{chen2021dynamic} uses the divergence of local models to the global model as the importance of clients, but the importance is used in client sampling instead of aggregating updates. A similar work in client sampling with priority is discussed in \cite{Goetz2019Active}, where a local loss value is used as client importance.
On the adaption and scheduling of the learning rates in FL: 
\cite{reddi2021adaptive} proposes adaptive optimization methods, such as Adagrad, Adam and YOGI, at the server side; 
\cite{charles2020outsized} proposes a heuristic rule to adjust server and local learning rates based on loss values.
All these existing works on local model update aggregation and learning rate adaption are all based on manually engineered strategies, which often require a great amount of efforts in both designing and experimentation. 

Our proposed algorithm belongs to the class of online meta-learning~\cite{Finn2019Online}.
In non-FL settings, ideas have be explored in the learning rate adaption and training example weighting:
gradient-based methods to learn the learning rate are proposed in \cite{gunes2018hypergrad}.
It is extended by \cite{Micaelli2020LongHorizons} from the next step objective to a long-horizon one. 
On training example weighting, Meta-weight-net~\cite{Shu2019MetaWeightNet} learns the weight of training examples to address the bias and noise in the training data.
In this thread, \cite{Zhang2021NoReward} proposes a method to learn without the need of a reward dataset and  
\cite{Xu2021famus} speeds up the meta-gradient computation with a faster layer-wise approximation.
In FL, \cite{chen2019federated} applies MAML~\cite{finn2017maml} and Meta-SGD~\cite{li2017metasgd} to learn the initial model weights for local training on each client.
However, the idea of using meta-learning to adapt the learning rate or client weights is yet to be explored in FL. 


\vspace{-2ex}
\section{Problem formulation}
CTR prediction can be formulated as a supervised learning problem: given the input feature and click/no click label pairs $\{x, y\}$,
the goal is to learn a function $h(x; w)$ parameterized by $w$, 
which minimizes the loss between the prediction and the ground truth label $\ell(h(x; w), y)$. At communication round $t$ of FL, a set of clients $S^t$ are selected for a participation, to whom a copy of 
the global model $w$ is distributed.
Each  client $k \in S^t$ trains the model using their local data $D_k$
to generate a local model $w_k^t$: 
$
w_k^t = \text{opt-local}(w^t; D_k).
$
$\text{opt-local}(\cdot)$ is the local optimization of $\ell(\cdot)$ 
initiated using weights $w^t$ on dataset $D_k$, which is often several epochs of SGD.
The clients send back to the server the model updates ${\Delta W}^t := \{ \Delta w_k^t, k \in S^t\}$, 
where ${\Delta w}_k^t := w_k^t - w^t$.
These local updates are then aggregated at the server to get the update on the global model 
$ 
\Delta w^t = \text{agg}({\Delta W}^t, Z^{t}).
$
where $Z^t := \{z_i, i \in S\}$ are the non-sensitive client attributes and $\text{agg}(\cdot)$ is the aggregation function which takes the local updates 
and the client attributes as input and outputs the aggregated update.
The aggregated update $\Delta w^t$ is then applied to the 
global model $w^t$ by performing one step of server optimization, i.e. $\text{opt-server}(\cdot)$, to get the new global model 
 $ w^{t + 1} = \text{opt-server}(w^t, \Delta w^t).$

Many existing FL algorithms can be formulated under this framework. A few examples are:
\noindent{$\blacktriangleright$~FedAvg \cite{mcmahan2017communication}}
$
\text{agg}({\Delta W}^t, Z^t) = \frac{1}{n^t} \sum_{k \in S} n^t_k \cdot \Delta w^t_k,
$
and\\$\text{opt-server}(w, \Delta w) = w^t + \Delta w^t,$
where $z_k^t = n_k, \text{ for } k \in S^t$, $n_k$ is the number of samples for client $k$, and 
$n^t = \sum_{k \in S^t} n_k$.

\noindent{$\blacktriangleright$~FedNova}\cite{Wang2020fednova}
$\text{agg}({\Delta W}^t, Z^t) =\big(\sum_{k \in S^t} \tau_k \cdot \frac{n_k}{n^t}\big) \cdot 
\sum_{k \in S} \frac{n_k}{n^t} \cdot \frac{1}{\tau_k} \cdot \Delta w_k^t,
$
where $z_k^t = [n_k, \tau_k]$, $\tau_k$ is the number of local gradient descent steps on 
client $k$. 
The function $\text{opt-server}(\cdot)$ is the same as that of FedAvg.

\noindent{$\blacktriangleright$~FedAdam and FedAdagrad}\cite{reddi2021adaptive}
These two methods have the same aggregation function as FedAvg. 
The server optimization method maintains first and second order momentum, $m^t$ and $M^t$:
$
  m^t = \beta_1 \cdot m^{t - 1} + (1 - \beta_1) \cdot \Delta w^t,
$
and 
$
 M^t = M^{t - 1} + (\Delta w^t)^2, \space \text{ for FedAdagrad},
$
$
M^t = \beta_2 \cdot M^{t - 1} + (1 - \beta_2) \cdot (\Delta w^t)^2, \text{   for FedAdam},
$
The optimization step is performed as: 
$
w^{t+1} = w^t + \gamma_s \cdot \frac{m^t}{\sqrt{M^t} + \epsilon},
$
where $\gamma_s, \beta_1, \beta_2 \text{ and } \epsilon$ are all hyper-parameters.
$\gamma_s$ is generally known as server learning rate.


\vspace{-2ex}
\section{Online Meta Learning for Local Update Aggregation}

To have a parameter-wise aggregation model, we will first define a partition of the weight indices $\mathcal{P}$, s.t. $\bigcup_{A \in \mathcal{P}} A$ is all the weight indices and 
$A_1 \cap A_2 = \varnothing, \text{for any pair of } A_1, A_2 \in \mathcal{P}, A_1 \neq A_2$. 
The most common partition is layerwise, where each $A \in \mathcal{P}$ is the set of weight indices of a network layer.
We will use $[\cdot]$ to denote the indexing operation on the weights.

\minisection{Meta model and its parameters} Our learnable aggregation model $\text{agg}(\cdot)$ is defined as:
\begin{equation*}\label{eq:agg_layer}
\begin{split}
    \Delta w^t[A] &= \text{agg}({\Delta W}^t[A], Z^t[A]; \theta[A]) := \theta_{s}[A] \cdot \sum_{k \in S^t} \alpha_k[A] \cdot \Delta w_k[A], \\
    & \text{ where } \alpha_k[A] = \text{softmax}(f_{\alpha}(z^t_k[A]; \theta_{\alpha}[A])),  
     \forall A \in \mathcal{P}.
\end{split}
\end{equation*}
$\theta = \{\theta_s, \theta_{\alpha}\}$ are the meta parameters to be learned. 
Each element of $\theta_s$ is between 0 and 1. 
It is the scaling on the server learning rate for step-size adaption during the learning process.
$f_{\alpha}(z; \theta_{\alpha})$ is the client weighting function parameterized by $\theta_{\alpha}$, which gives the importance of the clients
based on their non-sensitive attribute(s) $z$. 
$f_{\alpha}$ should be differentiable in $\theta_{\alpha}$. 
The selection of $f_{\alpha}$ and $z^t_k$ will be detailed in Sections~\ref{sec:alg} and~\ref{sec:experiments}.
The output of the meta model is the aggregated update $\Delta w^t[A], A \in \mathcal{P}$, which is applied to get the new weight: 

$w^{t+1}[A] = \text{opt-server}(w^t[A], \Delta w^t[A]), \forall A \in \mathcal{P}$.
This formulation allows the flexibility of aggregating different subsets of weight updates differently.
We will omit this indexing in the rest part of the paper for readability.

\minisection{Meta loss.}
Each local dataset $D_k$ is first split into 
the \textit{support} dataset $D^{(s)}_k$ for local training, 
and the \textit{query} dataset $D^{(q)}_k$ for meta loss evaluation.
The meta loss function is then defined in a \textit{delayed} way, i.e. the loss for the meta parameters at 
communication round $t - 1$, $\theta^{t - 1}$ will be evaluated at the round $t$:
\begin{equation}\label{eq:mlossa}
L^t(\theta^{t - 1})  = \sum_{k \in S^t}  L^t(\theta^{t - 1};  D_k^{(q)}),
\end{equation}
where each 

$L^t(\theta^{t - 1}; D_k^{(q)}) := $ $\sum_{\{x, y\} \in D_k^{(q)}}  \ell\big(h(x; w^{t}); y\big)$, abbreviated as $L^t_k(\theta^{t - 1})$.
The motivation is that, at the current round $t$, we evaluate the quality of our aggregation 
strategy performed at the previous round $t-1$ and adjust it accordingly. 
One advantage of (\ref{eq:mlossa}) that it is an unbiased estimation
of the training loss of $w^{t}$ from training examples on $S^t$.
Another advantage is that both its evaluation and optimization is done \textit{online}; this means that no extra datasets are needed nor are additional communication rounds required. 

\minisection{Meta optimization} 
We use gradient descent to optimize the meta parameters. The computation of the meta gradient is one of the main 
contributions of this paper. 
This section gives the description on the computation of $\frac{\partial L^t}{\partial \theta^{t - 1}} \big(\theta^{t - 1}\big)$. We will slightly abuse the notation by using $\theta^{t - 1}$ for both the variable and its value, which can be distinguished by the context.

Similar to the meta loss, its gradient is also computed in a \textit{delayed} way: the gradient of 
$\theta^{t - 1}$ is computed at round $t$. 
To do this, we will need to store the model weight, local model updates and clients' attributes of the previous round at the server,
so that we can assume that at round $t > 1$, we have 
$w^{t-1}, \Delta{w}^{t - 1}_k \text{ and } z^{t-1}_{k}, \forall k \in S^{t - 1}$ avaialble. 

By the chain rule, we have
\begin{equation}\label{eq:mgd}
\begin{split}
    \frac{\partial L^t}{\partial \theta^{t - 1}} \big(\theta^{t - 1}\big) = 
    \frac{\partial L^t} {\partial w^t}\big(w^t\big)
    \cdot \frac{\partial w^{t}}{\partial \theta^{t - 1}}\big(\theta^{t-1}\big).
\end{split}
\end{equation}
The first part of \eqref{eq:mgd} RHS, denoted as $g^t$, is the gradient of $L^t$ w.r.t $w^t$ evaluated at $w^t$ on the training data of the clients in $S^t$, i.e.

\begin{figure}[t]
    \centering
    \includegraphics[scale=0.5]{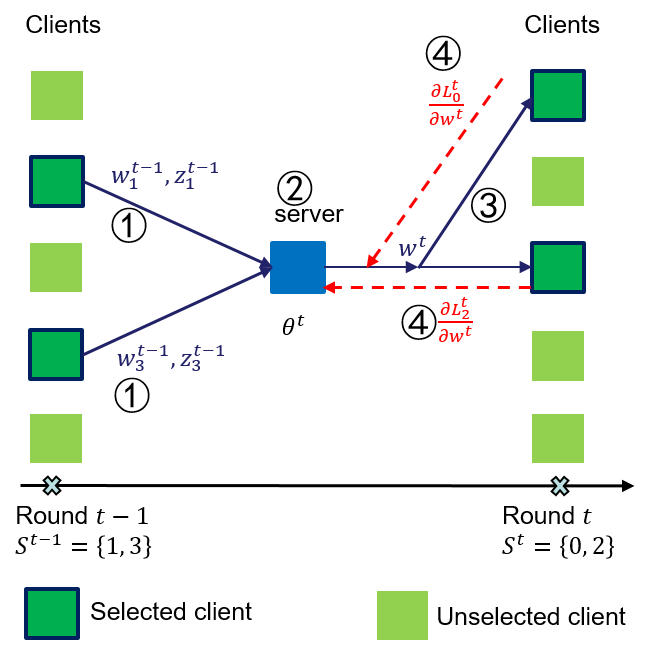}
    \caption{Overview of the gradient computation procedure in MetaUA: \textcircled{1} local model updates and attributes are sent from selected clients; \textcircled{2} local updates are aggregated using meta model; \textcircled{3} new model is distributed to the selected clients; 
    \textcircled{4} server receives the gradients from the selected clients and computes 
    the gradients for the meta-parameters. }
    \label{fig:metaua}
\end{figure}

$
 g^t = \sum_{k \in S^t }\underbrace{\frac{\partial L^t_k} {\partial w^t}\big(w^t\big)}_{g^t_k := }.
$
Each $g^t_k$ is evaluated locally on each client $k \in S^t$ as 
\begin{equation}\label{eq:gtk}
g^t_k = \sum_{x,y \in D_k^{(q)}} \frac{\partial \ell} {\partial w^t}\big(f(x, w^t), y\big),
\end{equation}
and then sent back to the server.
The second part of \eqref{eq:mgd} RHS is the Jacobian matrix  
${\partial w^{t}}/{\partial \theta^{t - 1}}$ evaluated at $\theta^{t-1}$.  
The dependency of $w^{t}$ on $\theta^{t - 1}$ is through $\Delta w^{t}$ as the composite of the two functions $\text{agg}(\cdot)$ and 
$\text{opt-server}(\cdot)$. 
Both of these two functions are differentiable, so we can also apply the chain rule: 
$
    \frac{\partial w^{t}}{\partial \theta^{t - 1}} = \frac{\partial w^{t}}{\partial \Delta w^{t-1}} \cdot 
    \frac{\partial \Delta w^{t-1}} {\partial \theta^{t - 1}}.
$
In the implementation, there is no need to compute the full Jacobian matrices.
Instead, since $g^t$ is not a function, we move it into the second paritial derivative in Eq.~\eqref{eq:mgd}, which becomes:
\begin{equation}\label{eq:vjp}
 \frac{\partial L^t}{\partial \theta^{t - 1}}\big(\theta^{t-1}\big) = \frac{\partial (g^t \cdot w^t)}{\partial \theta^{t - 1}}
 \big(\theta^{t-1}\big) .
\end{equation}
The inner product $g^t\cdot w^t$ is a scalar-output linear function of $w^t$. Its gradient w.r.t $\theta^{t - 1}$ can 
easily be computed through back propagation provided by most automatic differentiation packages. The procedure of gradient 
computation in MetaUA in two FL rounds is presented in Fig.~\ref{fig:metaua}.

\minisection{MetaUA Algorithm} 
\label{sec:alg}
The proposed algorithm of Online \textbf{Meta}-Learning~for \textbf{U}pdate \textbf{A}gg\-regation (\textbf{MetaUA}) is given in Algorithm~\ref{alg}, where all computations follow previously described derivations.
The key difference between it and the other FL algorithms is that, at each communication round, 
the meta parameters will be updated by a step of gradient descent, which are then used to aggregate local updates to form a new version of the server's model.

\SetAlCapNameFnt{\footnotesize}
\SetAlCapFnt{\footnotesize}
\begin{algorithm}
\footnotesize
\DontPrintSemicolon
    
    \KwInput{ Randomly initialized weight $w^1$}
    \KwOutput{ $w$ after the learning is complete}
    \KwData{ An universe of clients $I$, for each client $k \in I$, 
    an associated local dataset $D_k = D_k^{(s)} \cup D_k^{(q)}$}
        
    \For{communication round t = 1, 2, \dots, T} {
        Randomly select a set of clients $S^t$ \\
        \For { each client $k$ in $S^t$ }{
            $\Delta w^t_k \leftarrow \text{opt-local}(w^t, D_k^{(s)}) - w^t$  \tcp*{local training}
            compute $g^t_k$ with equation (\ref{eq:gtk}) \\ 
            set client attributes to $z^t_k$ \\
        }

        \If{  $t > 1$ } {
            \tcc{backward pass of update aggregation}
            $g^t \leftarrow \sum_{k \in S^t} g^t_k$ \\
            compute $\frac{\partial L^t}{\partial \theta^{t - 1}}\big(\theta^{t - 1}\big)$ with equation (\ref{eq:vjp}) \\
            $\theta^t \leftarrow \theta^{t - 1} - \gamma_{meta} \cdot \frac{\partial L^t}{\partial \theta^{t - 1}}\big(\theta^{t - 1}\big) $
        }
        
        \tcc{forward pass of update aggregation}
        $\Delta w^t \leftarrow \text{agg}(\Delta W^t, Z^t; \theta^t)$  \\ 
        $w^{t + 1} \leftarrow \text{opt-server}(w^t, \Delta w^t)$  
    }
\caption{Online Meta-Learning for Update Aggregation (\textbf{MetaUA})}
\label{alg}
\end{algorithm}

\vspace{-1.5ex}
\paragraph{Server optimizer}
Our algorithm is flexible in the selection of the server optimizer.
We run experiments on SGD, FedAdagrad and FedAdam, in which FedAdagrad has the best results.
We thus choose to use FedAdagrad in all the experiments in this paper.

\vspace{-1.5ex}
\paragraph{Client weighting model}
\label{sec:metaua_z}
The main consideration in choosing $f_{\alpha}$ is that it will run for all the selected clients in both forward 
and backward pass, therefore, the use of a simple architecture to keep the additional computations
small should be favored. 
In our experiments, we found that using a simple linear model for $f_{\alpha}$ is enough to get
good results.

For client attributes, a good candidate should: 
1) not reveal user privacy; 
2) capture the client heterogeneity;
3) introduce little or no extra computation.
Inspired by \cite{Shu2019MetaWeightNet} and \cite{Goetz2019Active}, we choose the local loss 
$z^t_k = \frac{1}{n_k}\sum_{\{x, y\} \in D_k} \ell(f(x; w^t); y)$.
Intuitively, the local loss measures how well the current model fits the client's local data. 
A higher local loss than others could be caused by either system or data heterogeneity.
Some possible reasons are:
the client has lagged behind others in learning because it has not actively participated in FL, 
or less local training steps have been performed on it because it has less powerful hardware, 
or the client has examples that the current model fits poorly, 
which could be examples novel to the model, ambiguous examples near the decision boundary or mislabelled  examples.
We have also considered including: number of samples,norm of the weight updates, number of unique features, etc. 
Experimental results are provided in section~\ref{sec:experiments}.

$\gamma_{meta}$ is the learning rate on the meta-parameters. We find that our method is generally robust to it. In our experiments, we fix it to be $0.1$, which has reasonably good results under all the settings.

\vspace{-2ex}
\section{Cost Analysis}
Compared to other FL methods there are some extra costs incurred from MetaUA at each training round. 

\minisection{Communication} 
To compute $g^t$, we need to transfer the gradient of the local loss w.r.t the model parameter to the server for all the selected clients, i.e. $g^t_k, k \in S^t$.
This will double the communication cost from the clients to the server.
A few methods can be explored to reduce the communication: compressing the model gradient before sending or sub-sampling the clients in gradient evaluation.

\minisection{Computation} On the client side, the extra computation needed is for $g^t_k, k \in S^t$. 
This requires one additional pass over the local data. We consider the increase in computation to be marginal, as each selected client already have to perform several epochs of training in FL round.
On the server side, an extra computation is needed for: the meta aggregation operator, the opt-server, the inner product in \eqref{eq:vjp} and a backward pass on them for the gradient computation.
The aggregation operator involves shallow networks to be applied as many times as the number of selected clients in an FL round, while other operations are just a constant number of simple operations. 
The impact of this extra computation cost on the server side is not significant, because the server is usually a high-performance cluster.

\minisection{Storage} Since our meta gradient evaluation is delayed, we need to store the model updates received from the selected clients for a single communication round. In practice, the user population is quite large, so additional memory resources have to be allocated for that.

\vspace{-2ex}
\section{Experiments}
\label{sec:experiments}
We evaluated a proposed method for a click-through rate (CTR) prediction task in FL setting that can be used in 
real-world scenario. A series of experiments are designed and conducted to present methods properties.

\minisection{Datasets}
In our experiments we use four well-know datasets: MovieLens-1M, Tmall, Yelp, and Amazon-Cds. 
Their main statistics are presented in Table~\ref{tab:datasets}.
Input features can be related to: user, item or context of the event. Data of each user is taken as
a client's partition in all federated learning experiments. The last 10\% of client's data (based on a timestamp) 
is taken for a validation.

\begin{table}[tb]
\centering
\resizebox{0.97\columnwidth}{!}{%
\begin{tabular}{lrrrrrr}
\toprule
Dataset &  \# Examples & \# Features  &  Vocab Size  & \# Users &  \# Items &  Density \\
\midrule
MovieLens-1M &  739012 &  7 & 13196 & 6040 & 3668 & 3.34 \\
Tmall & 1899378 & 4 & 44239 &  22284 & 17705 & 0.48 \\
Yelp & 530124 & 9 &  34462 & 22128 & 12232 &  0.20\\
Amazon-CDs & 890824 & 3 &  31985 & 15592 & 16184 &   0.35\\
\bottomrule
\end{tabular}
}
\caption{Dataset statistics}
\label{tab:datasets}
\end{table}

\minisection{Implementation Details} In all experiments we use DCNv2~\cite{wang2021dcn} prediction model with binary
cross entropy as a loss function and dimension for all embeddings set to $4$. For a
central training experiments Adam optimizer is used for ten epochs with
$lr=0.0001$, weight decay $0.0001$, batch size of $256$. In FL settings we use SGD as a local optimizer 
with $lr=0.01$ with batch size of $15$ and three epochs done in each of 200 communication rounds. In proposed method, 
meta-parameters are optimized with SGD with $lr=2.0$. We use layer-wise approach where each parts of the network
have dedicated meta-parameters. Implementation is done using PyTorch library.

\begin{figure*}[t]
\centering
\includegraphics[width=0.8\textwidth,keepaspectratio]{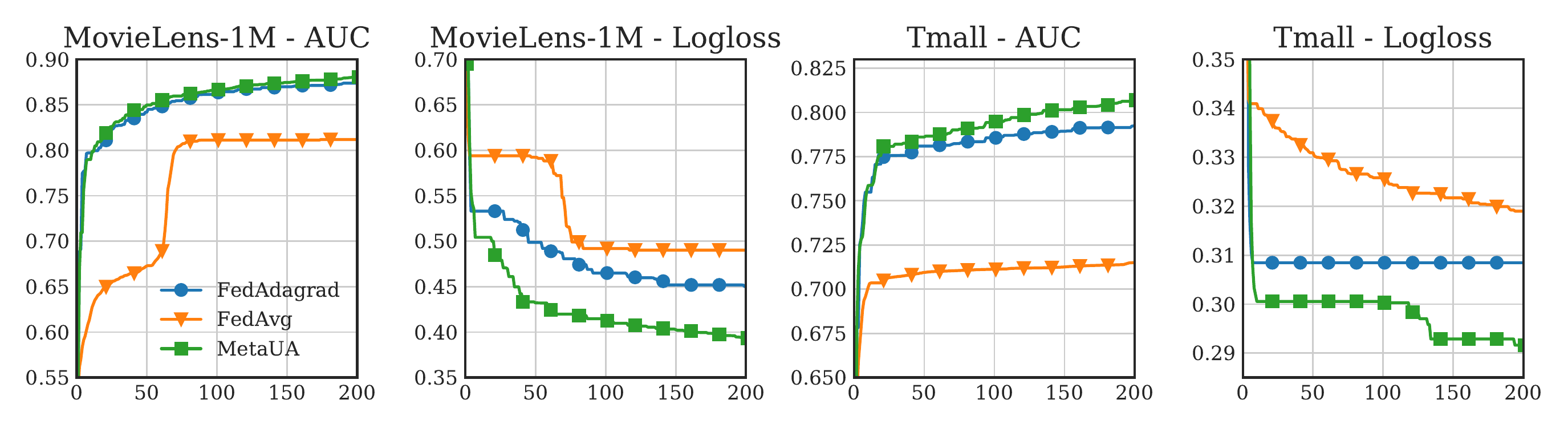}
\vspace*{-3ex}
\caption{AUC and Logloss values during the federated train with methods FedAvg, FedAdaptive and MetaUA for
MovieLens-1M and Tmall datasets.}
\label{fig:cmp_plots}
\end{figure*}

\minisection{Comparison with other methods} The results for comparison of MetaUA with FedAvg and FedAdagrad are presented in Table~\ref{tab:methods_cmp}.
Both adaptive methods are better than simple FedAvg, they converge faster, and the training process is more stable.
MetaUA got slightly better final AUC than FedAdagrad or very competitive (AmazonCDs worse only by $0.007$). 
However, if we consider for those results logloss value during the train the difference is big. On average MetaUA
gets $8.5\%$ better logloss. This margin is noticeable in FL training curves 
in Figure~\ref{fig:cmp_plots}.

\begin{table}[tb]
\centering

\resizebox{\columnwidth}{!}{%
    \begin{tabular}{ll|rrr|rrr}
\toprule
     & & \multicolumn{3}{c}{AUC} & \multicolumn{3}{c}{Logloss} \\
 Dataset & Round &  FedAvg & FedAdagrad & MetaUA &  FedAvg & FedAdagrad & MetaUA \\
\bottomrule
\toprule
ML-1M & 20  &  0.648 &      0.809 &  \textbf{0.814} &   0.594 &      0.533 &  \textbf{0.500} \\
     & 50  &  0.673 &      0.842 &  \textbf{0.850} &   0.592 &      0.499 &  \textbf{0.432} \\
     & 100 &  0.811 &      0.864 &  \textbf{0.867} &   0.492 &      0.465 &  \textbf{0.413} \\
     & 150 &  0.811 &      0.870 &  \textbf{0.875} &   0.490 &      0.452 &  \textbf{0.403} \\
     & 200 &  0.812 &      0.874 &  \textbf{0.880} &   0.490 &      0.450 &  \textbf{0.393} \\
\midrule
Tmall & 20  &  0.704 &      0.775 &  \textbf{0.779} &   0.338 &      0.308 &  \textbf{0.301} \\
     & 50  &  0.709 &      0.781 &  \textbf{0.786} &   0.331 &      0.308 &  \textbf{0.301} \\
     & 100 &  0.711 &      0.786 &  \textbf{0.794} &   0.326 &      0.308 &  \textbf{0.300} \\
     & 150 &  0.713 &      0.789 &  \textbf{0.802} &   0.322 &      0.308 &  \textbf{0.293} \\
     & 200 &  0.715 &      0.792 &  \textbf{0.807} &   0.319 &      0.308 &  \textbf{0.292} \\
\midrule
Amzn & 20  &  0.588 &      0.605 &  \textbf{0.640} &   0.693 &      0.670 &  \textbf{0.662} \\
     & 50  &  0.599 &      0.668 &  \textbf{0.808} &   0.693 &      0.670 &  \textbf{0.541} \\
     & 100 &  0.600 &      0.899 &  \textbf{0.919} &   0.693 &      0.434 &  \textbf{0.335} \\
     & 150 &  0.601 &      \textbf{0.940} &  0.937 &   0.693 &      0.320 &  \textbf{0.286} \\
     & 200 &  0.604 &      \textbf{0.951} &  0.944 &   0.693 &      0.279 &  \textbf{0.269} \\
\midrule
Yelp & 20  &  0.579 &      \textbf{0.709} &  0.708 &   0.692 &      0.653 &  \textbf{0.631} \\
     & 50  &  0.595 &      0.745 &  \textbf{0.758} &   0.686 &      0.653 &  \textbf{0.596} \\
     & 100 &  0.603 &      0.762 &  \textbf{0.781} &   0.682 &      0.652 &  \textbf{0.576} \\
     & 150 &  0.607 &      0.770 &  \textbf{0.794} &   0.680 &      0.647 &  \textbf{0.564} \\
     & 200 &  0.611 &      0.774 &  \textbf{0.801} &   0.678 &      0.641 &  \textbf{0.559} \\
\bottomrule
\end{tabular}

}
\caption{Performance evaluation of FedAvg, FedAdagrad and MetaUA over different datasets. For a comparison
logloss and AUC are presented for different rounds of federated learning. Best values are in bold.}
\label{tab:methods_cmp}
\end{table}

\begin{figure}[tb]
\centering
\subfigure{\includegraphics[width=0.2\textwidth,keepaspectratio]{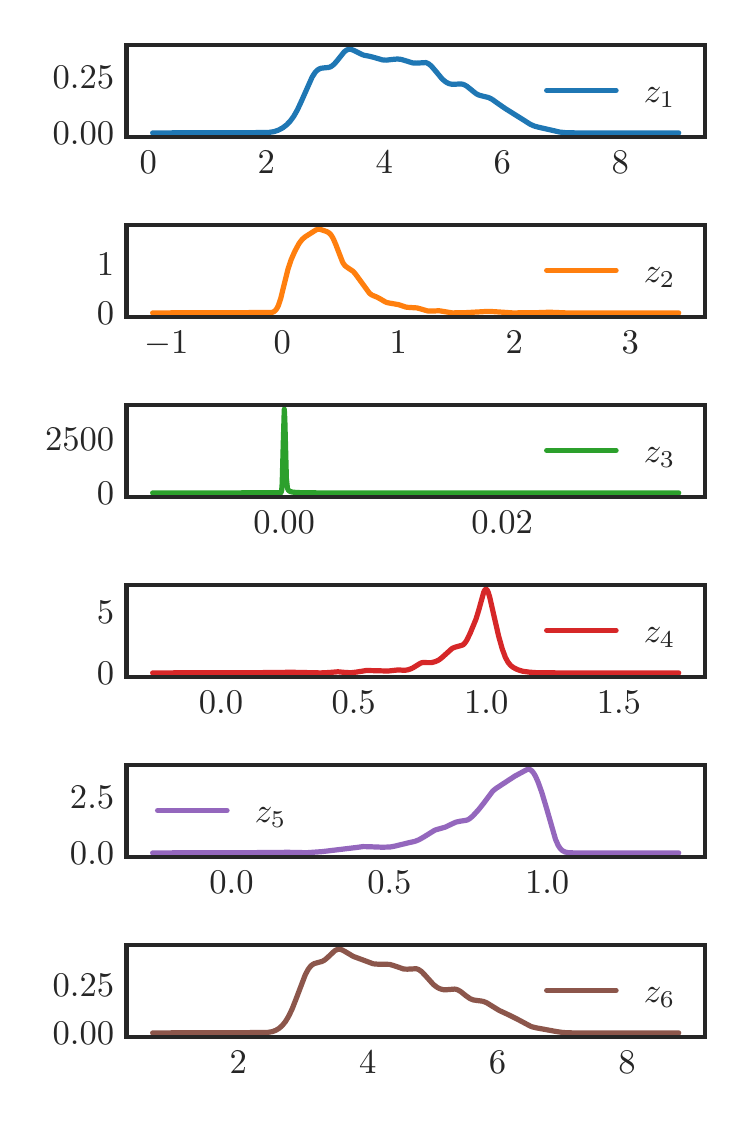}}
\subfigure{\includegraphics[width=0.2\textwidth,keepaspectratio]{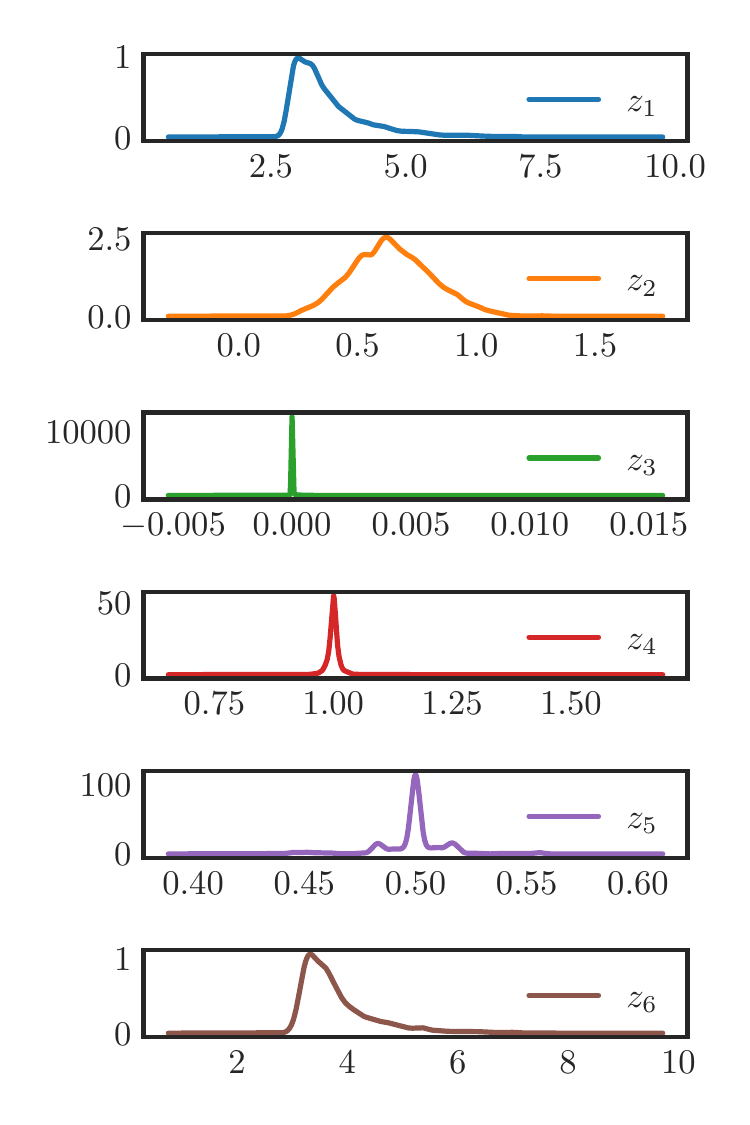}}
\caption{Distribution of $Z$ variables for datasets Movielens-1M (left) and Amazon-CDs (right). 
Snapshot of sampled values are taken during round number 20.}
\label{fig:metaua_z_dist_1}
\end{figure}

\minisection{Selection of client's attributes}
We evaluated a few different information that can be shared to a server from clients during a FL training. 
Following desired characteristic described in section~\ref{sec:metaua_z} we used: 
$z_1$ - number of samples~\cite{mcmahan2017communication}, 
$z_2$ - local loss value~\cite{Shu2019MetaWeightNet},
$z_3$ - gradient norm,
$z_4$ - ratio of loss value after local SGD train to local loss,
$z_5$ - positive class probability, 
and $z_6$ - number of unique features in local dataset~\cite{Wistuba2016TwoStageTS,Wistuba2017ScalableGP,Jomaa2019HypRLH}.
Each of the variables grasps a different aspect of a client's local dataset and local training process during 
the FL communication round. Sampled clients $Z$ distributions during FL train for MoveieLens-1M and Amazon-CDs datasets 
are presented in Figure~\ref{fig:metaua_z_dist_1}.
The smallest variation presents $z_3$. However, this is the only one dependent on FL round as well as the
layer of the network (computed layer-wise, on plots we see aggregated values).

Table~\ref{tab:metaua_z_amz} present the results for Amazon-CDs dataset.
The best final logloss value is achieved only using $z_2$ (local loss), which happens to be 
as well a good proxy for client importance in previous work~\cite{Shu2019MetaWeightNet}. Up to some point in 
FL training, using only $\theta_s$ is also beneficial. $z_6$ (unique features num.) and $z_2$ are second good choices. 
Surprisingly, taking all variables does not yield the best results. For evaluated datasets it seems like 
simpler choices and models works better with an online meta-learning. In our experiments we evaluated 
different models, i.e.~MLP network for $f_{\alpha}(Z; \theta_{\alpha})$. However, the results for simple linear
model were better. 

\begin{table}[tb]
\centering
\resizebox{\columnwidth}{!}{%
    \begin{tabular}{lrrrrr|rrrrr}
\toprule
{} & \multicolumn{5}{c}{AUC} & \multicolumn{5}{c}{Logloss} \\
rounds &    20  &    50  &    100 &    150 &    200 &     20  &    50  &    100 &    150 &    200 \\
method &        &        &        &        &        &         &        &        &        &        \\
\midrule
None   &  0.640 &  0.761 &  0.911 &  0.933 &  0.943 &   \textbf{0.661} &  0.606 &  0.362 &  0.298 &  0.269 \\
$z_1$  &  0.700 &  0.897 &  0.936 &  0.949 &  0.956 &   0.685 &  0.399 &  0.300 &  0.261 &  0.241 \\
$z_2$  &  0.629 &  0.637 &  0.887 &  0.944 &  \textbf{0.964} &   0.667 &  0.667 &  0.405 &  0.275 &  \textbf{0.206} \\
$z_3$  &  0.595 &  0.690 &  0.825 &  0.871 &  0.894 &   0.693 &  0.693 &  0.676 &  0.573 &  0.498 \\
$z_4$  &  0.611 &  0.641 &  0.815 &  0.881 &  0.922 &   0.679 &  0.679 &  0.540 &  0.419 &  0.335 \\
$z_5$  &  0.637 &  0.650 &  0.887 &  0.926 &  0.944 &   0.666 &  0.666 &  0.404 &  0.320 &  0.275 \\
$z_6$  &  \textbf{0.712} &  \textbf{0.899} &  \textbf{0.937} &  \textbf{0.950} &  0.957 &   0.684 &  \textbf{0.391} &  \textbf{0.296} &  \textbf{0.257} &  0.239 \\
All    &  0.592 &  0.728 &  0.808 &  0.845 &  0.864 &   0.691 &  0.664 &  0.664 &  0.664 &  0.649 \\
\bottomrule
\end{tabular}

}
\caption{Performance of MetaUA evaluated with different set of client's attributes $Z$ on Amazon-CDs dataset.}
\label{tab:metaua_z_amz}
\end{table}

\minisection{Ablation study} In order to evaluate contribution of different elements of MetaUA method we performed ablation study where each 
part of the approach is disabled. Four versions are evaluated in very same setting: only server learning rate 
meta-learning, client weighting with meta-learning, both at the same time and no adaptation at all.
Comparison is presented in Table~\ref{tab:ablation} for MovieLens-1M dataset. 
While the model is optimized toward the best logloss value, here application of whole MetaUA gives the best advantage. 
For a longer FL training period, at round 400, adaptation of server learning rate and client weighting yield
similar outcome of $0.413$. Only the application both at once lower the results to $0.382$. AUC seems also 
beneficial when all elements of MetaUA are used. 

\begin{table}[tb]
\centering
\resizebox{\columnwidth}{!}{%
    \begin{tabular}{l|rrrr|rrrr}
\toprule
{} & \multicolumn{4}{c}{\textbf{AUC}} & \multicolumn{4}{c}{\textbf{Logloss}} \\
FL round &    50  &    100 &    200 &    400 &     50  &    100 &    200 &    400 \\
\midrule

\midrule
no adjustment                &  0.841 &  0.864 &  0.878 &  0.882 &   0.502 &  0.461 &  0.436 &  0.429 \\
client weighting             &  0.844 &  0.862 &  \textbf{0.880} &  0.885 &   0.454 &  0.439 &  0.417 &  0.413 \\
server lr                    &  0.846 &  0.865 &  0.877 &  0.881 &   0.484 &  0.441 &  0.415 &  0.413 \\
client weighting + server lr &  \textbf{0.850} &  \textbf{0.867} &  \textbf{0.880} &  \textbf{0.889} &   \textbf{0.432} &  \textbf{0.413} &  \textbf{0.393} &  \textbf{0.382} \\

\bottomrule
\end{tabular}

}
\caption{Ablation study}
\label{tab:ablation}
\end{table}

\begin{figure}[tb]
\centering
\subfigure{\includegraphics[width=0.8\linewidth,keepaspectratio]{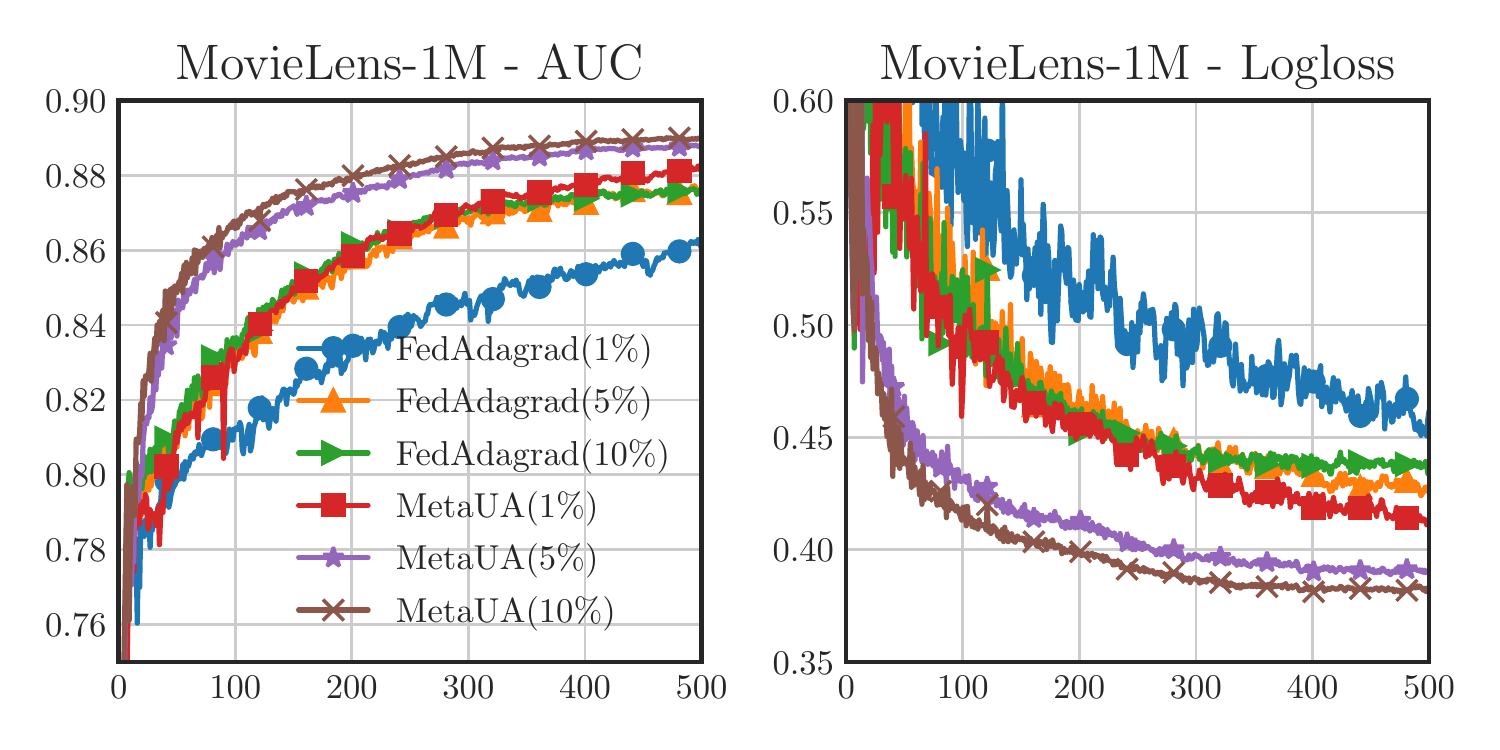}}\\[-3ex]
\subfigure{\includegraphics[width=0.8\linewidth,keepaspectratio]{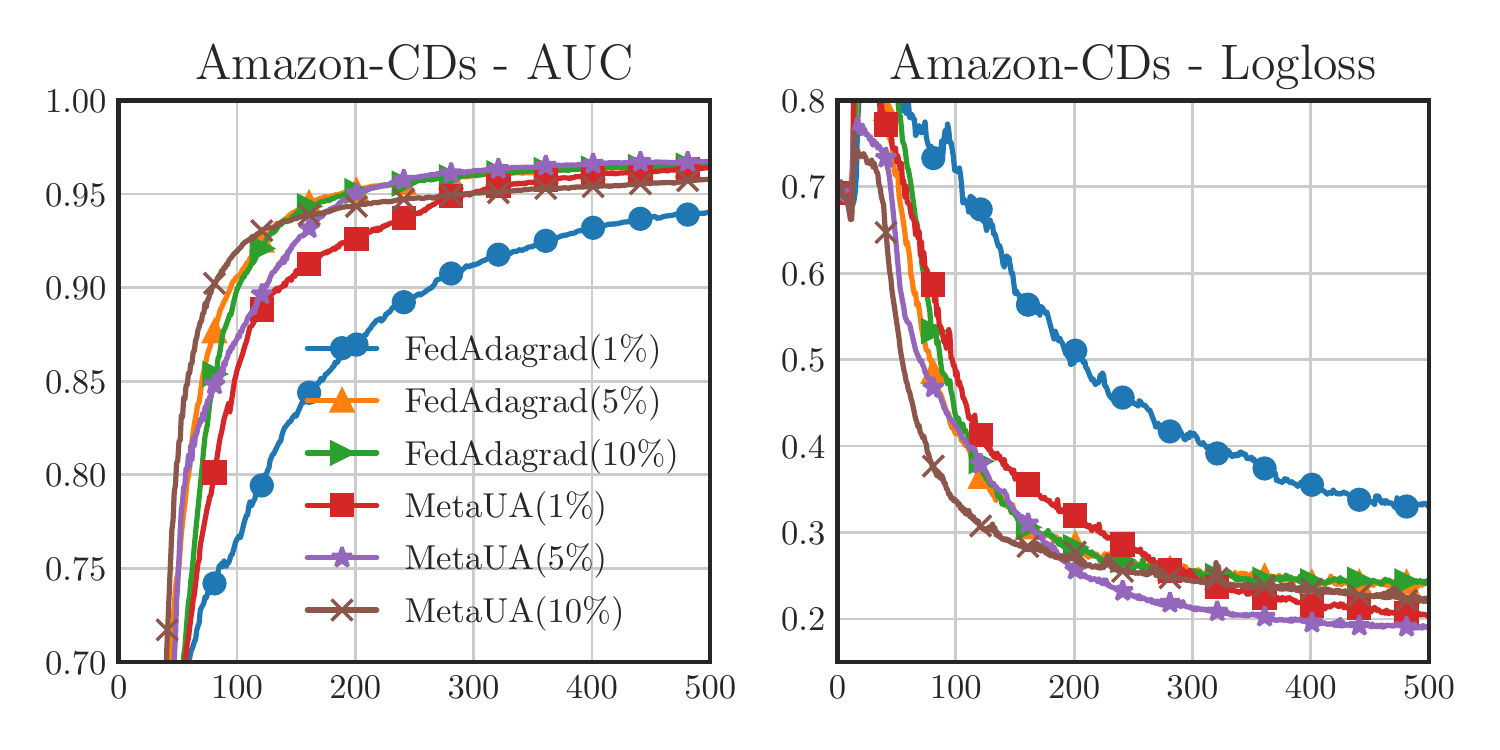}}
\vspace*{-1.7em}
\caption{Comparison of MetaUA and FedAdagrad training curves when different fraction of clients is used during FL rounds.}
\label{fig:fl_frac_1}
\end{figure}

\minisection{Different fraction of clients} To present robustness of MetaUA towards different fraction of client sampled during FL training rounds, we 
lowered this value to $5\%$ and $1\%$. Comparison with FedAdagrad is presented in Figure~\ref{fig:fl_frac_1}.
We presented a longer training as with fewer clients during training more rounds can be necessary for the convergence. 
While fraction get smaller, MetaUA still wins with FedAdagrad with a big margin. For MovieLens-1M, the difference
is clear. Only FedAdagrad with fraction $5\%$ outperforms $10\%$ with more than 350 rounds. For Amazon-CDs 
the best at final round got MetaUA with $5\%$ clients sampled in each round.    

\begin{figure}[tb]
\centering
\includegraphics[width=0.8\linewidth,keepaspectratio]{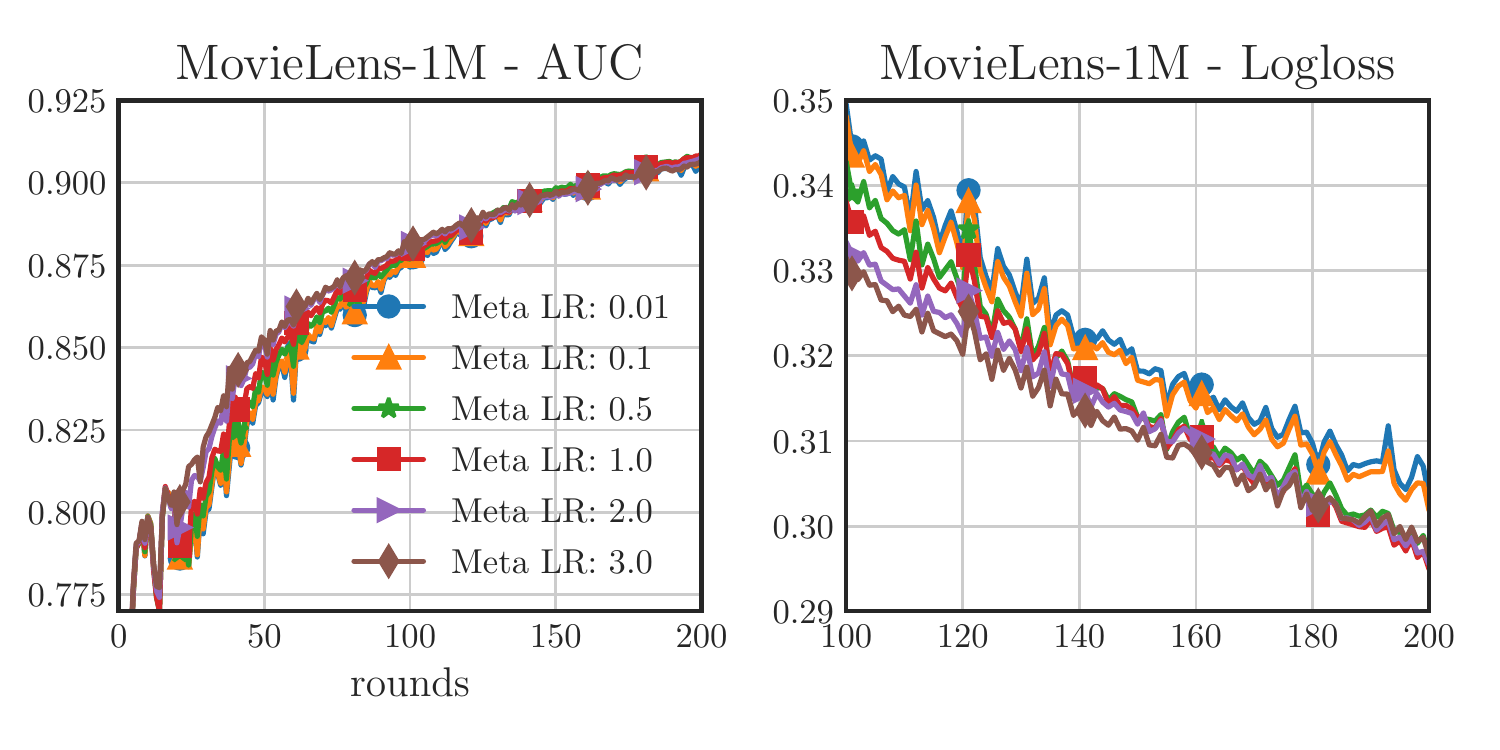}
\vspace*{-1.7em}
\caption{AUC and Logloss values during FL training on MovieLens-1M dataset for a various meta-learning rate value.
Logloss (right) plot is zoomed-in after 100 rounds to better present the difference.
}
\label{fig:metaua_lr_ml1m}
\end{figure}

\minisection{Robusness to hyper-params - selection of meta lr}
MetaUA is helping to solve its main task -- adaptive optimization during FL on the server side.
However, as any other meta-learning method, it introduces its own hyper-parameters at the meta-level.
As long as those are easy to find the whole process is applicable. For MetaUA the meta-learning rate is the one
most crucial. In all experiments we used fixed value of $2.0$, without doing extensive hyper-parameter search
each time. Thus, in Figure~\ref{fig:metaua_lr_ml1m} 
we can see the performance for different values of the meta-learning rate on MovieLens-1M dataset in order
to present the robustness of our approach. It is easy to find the best performing meta-learning. Here, 
for values bigger than one almost the same logloss at round 200 is received.

\vspace{-2ex}
\section{Conclusions}
This paper presents a method of learning adaptive models for client weighting and server learning rate adaption in the aggregation of local model updates for federated deep CTR models.
In our experiments on public benchmarks, our method shows better performance compared to the state-of-the-art methods. The method is adjusted in an online way fitted in FL communication rounds, without a need of additional heavy communication from the clients.
As future work, one important direction it to improve the privacy protection of our method, for example, by introducing a random noise to both model updates and gradients communicated back to the server.
Another extension of a proposed MetaUA method is to learn the sample weighting for local trains in the FL settings. 

{\footnotesize
\bibliographystyle{ACM-Reference-Format}
\bibliography{refs}
}

\end{document}